\documentclass[12pt]{article}
\usepackage[english]{babel}
\usepackage[latin1]{inputenc}
\usepackage{amsfonts,amssymb,amsmath, epsfig}
\usepackage{color,graphicx,graphics,psfrag}
\usepackage{amsmath,amstext,amssymb,amsfonts, amscd}
\usepackage{hyperref}          

\def\Box{\leavevmode\vbox{\hrule
     \hbox{\vrule\kern4pt\vbox{\kern4pt}%
           \vrule}\hrule}}

\newcounter{appendix}
\def\appendix{\advance\c@appendix by 1
   \def\thesection{\Alph{section}}
   \ifnum\c@appendix=1 \setcounter{section}{-1} \fi
   \@startsection {section}{1}{\z@}{-3.5ex plus -1ex minus 
   -.2ex}{2.3ex plus .2ex}{\Large\bf}}

\def\paragraph#1{{\bf #1\ }}

\newtheorem{lemma}{Lemma}[section]  

\newtheorem{theorem}[lemma]{Theorem}

\title{Macroscopic models of collective motion and self-organization} 
\author{Pierre Degond$^{1,2}$, Amic Frouvelle$^3$, Jian-Guo Liu$^4$, \\Sebastien Motsch$^5$, Laurent Navoret$^6$} 

\date{} 
\begin{document}

\maketitle

\vspace{0.5 cm}

\begin{center}
1-Université de Toulouse; UPS, INSA, UT1, UTM ;\\ 
Institut de Mathématiques de Toulouse ; \\
F-31062 Toulouse, France. \\
$\mbox{}$ \\
2-CNRS; Institut de Mathématiques de Toulouse UMR 5219 ;\\ 
F-31062 Toulouse, France.\\
email: pierre.degond@math.univ-toulouse.fr
\end{center}

\begin{center}
3- CEREMADE, UMR CNRS 7534\\
Université Paris-Dauphine\\
75775 Paris Cedex 16, France\\
email: frouvelle@ceremade.dauphine.fr
\end{center}

\begin{center}
4- Department of Physics and Department of Mathematics\\
Duke University\\
Durham, NC 27708, USA\\
email: jliu@phy.duke.edu
\end{center}

\begin{center}
5- Center for Scientific Computation and Mathematical Modeling (CSCAMM)\\
University of Maryland\\
College Park, MD 20742, USA\\
email: smotsch@cscamm.umd.edu
\end{center}

\begin{center}
6- Institut de Recherche Mathématique Avancée de Strasbourg \\
CNRS UMR 7501 and Université de Strasbourg \\
7 rue René Descartes, 67084 Strasbourg Cedex, France. \\
email: laurent.navoret@math.unistra.fr
\end{center} 

\newpage
\begin{abstract}
In this paper, we review recent developments on the derivation and properties of macroscopic models of collective motion and self-organization. The starting point is a model of self-propelled particles interacting with its neighbors through alignment. We successively derive a mean-field model and its hydrodynamic limit. The resulting macroscopic model is the Self-Organized Hydrodynamics (SOH). We review the available existence results and known properties of the SOH model and discuss it in view of its  possible extensions to other kinds of collective motion. 
\end{abstract}

\medskip
\noindent
{\bf Acknowledgements:} This work has been supported by the french 'Agence Nationale pour la Recherche (ANR)' in the frame of the contract 'MOTIMO' (ANR-11-MONU-009-01).

\medskip
\noindent
{\bf Key words: } Individual-Based Models, self-propelled particles, self-alignment, Vicsek model, mean-field kinetic model, Fokker-Planck equation, macroscopic limit, von Mises-Fisher distribution, self-organized hydrodynamics

\medskip
\noindent
{\bf AMS Subject classification: } 35L60, 35K55, 35Q80, 82C05, 82C22, 82C70, 92D50.
\vskip 0.4cm

\setcounter{equation}{0}
\section{Collective dynamics and self-organization}
\label{intro}

Many different kinds of interacting agent systems can be observed in nature, such as bird flocks, fish schools, insect swarms, etc. They provide fascinating examples of self-organizing systems which are able to produce large scale stable coherent structures: a typical example of such a structure is a social insect nest (such as a termite nest). The structure scale exceeds the insect typical size by several orders of magnitude, and no agent in the community has the cognitive capacity of planning it. Therefore, it emerges as a product of the local interaction between the agents (here their ability to manipulate mud bullets in response to chemical signals deposited by the other agents), without the intervention of a leader \cite{Khuong_etal_ECAL11}. Self-organization is ubiquitous and can be observed in the inanimate world (see e.g. the formation of galaxies, crystals, tornadoes), as well as in the living world (see examples above) and the social worlds (e.g. in traffic, crowds, opinion formation, finance, etc.). It appears at so many different scales that one can wonder whether the fate of the universe is a journey towards ultimate disorder, as pictured by Boltzmann in the concept of entropy. The concept of evolution to disorder and self-organization are somehow contradictory. The biologist J. Monod realized that these two observations are not easily reconciled and tried to do so in his famous essay 'Chance and Necessity' \cite{Monod_71}.

Another concept related to self-organizing systems is that of criticality. Indeed, such systems exhibit phase transitions between a disordered state and self-organized ones (see a review in \cite{Vicsek_Zafeiris_PhysRep12}). The 'thermodynamic variable' which induces phase transitions is often related to the intensity of the noise undergone by the agents in their motion. This provides an easy analogy to the temperature in classical thermodynamics. However, most often, the density or size of the system is another variable inducing phase transitions and more surprisingly, order appears at large densities. This is somewhat in contradiction with many physical systems in which high densities are associated to high temperature. The quantification of the amount of order exhibited by the system is done by means of a so-called order parameter, which usually ranges between $0$ and $1$ and which increases with the amount of order in the system. 

When the density or noise intensity are varied, the order parameter exhibits a behavior which is similar to those of phase transitions in physics (see e.g. \cite{Degond_etal_note_submitted13, Degond_etal_preprint13}). Two typical behaviors can be observed. The first one is that of first-order (or discontinuous) phase transitions: in this case, there is a parameter range of metastability in which the disordered state and the ordered one coexist. The transition from disorder to order or vice-versa results in a jump of the order parameter. Additionally, the jumps in either ways do not occur for the same value of the density (or noise), leading to a hysteresis behavior. The second behavior is that of second-order (or continuous) phase transitions. In this case, the transition from disorder to order results in a continuous (but singular in its derivative) change of the order parameter. In the region of abrupt change, the system is said to be in a critical state. 

Critical phenomena associated to self-organizing systems are grouped into the category of 'Self-Organized Criticality' \cite{Bak_etal_PRL87}. By contrast to physical systems where critical states are reached for very particular combination of the thermodynamic parameters, the critical states of self-organizing systems appear like attractors of the dynamics, leading to the belief that most of living or social systems operate at the critical state. 

We refer the reader to \cite{Vicsek_Zafeiris_PhysRep12} for a review of these subjects. The modeling of such self-organizing systems offers a number of new mathematical challenges, some of which are reviewed in the present work. One of the most important challenges, as discussed in  \cite{Vicsek_Zafeiris_PhysRep12} is the lack of conservation relations, which are the corner-stones of the macroscopic theory of large particle systems in physics. Another challenge is related to the possible breakdown of the propagation of chaos property, which is another corner stone of the statistical mechanics theory of large particle systems. 

As discussed above, these systems exhibit phase transitions. We have emphasized the importance of symmetry-breaking phase transitions, which bring the system from disordered to ordered states. The broken symmetry is most of the time the rotational invariance symmetry \cite{Degond_etal_note_submitted13, Degond_etal_preprint13}, as collective motion exhibits coherent motion in one preferred direction. Most often, this preferred direction is random and emerges spontaneously from the breaking of the rotational invariance of the fully disordered or isotropic state. But another kind of phase transition occurs, and is related to the finite size of the particles. It is the transition from the unjammed state (where the particles have room to move independently one from each other) to the jammed state, where they are at contact to each other, and any motion of one of them induces correlated motion of the neighboring particles, sometimes over very large distances. This jamming transition also calls for specific mathematical techniques \cite{Degond_etal_JSP10}. 

In this work, we will discuss the first challenge, i.e. how to obtain macroscopic models for systems which lack conservation relations. We will also briefly review questions related to symmetry-breaking phase transitions (a more detailed review can be found in \cite{Degond_etal_note_submitted13}). We will not discuss the breakdown of propagation of chaos, and refer the interested reader to \cite{Carlen_etal_PhysicaD13, Carlen_etal_M3ASD13}. We will also leave the discussion of the jamming transition for a future review. We just mention that this question has been theoretically discussed in relation to traffic models in \cite{Appert-Rolland_etal_NHM11, Berthelin_etal_ARMA08, Berthelin_etal_M3AS08, Degond_Delitala_KRM08} and to herding in \cite{Degond_etal_JSP10}. It has been numerically investigated in \cite{Degond_Hua_JCP13, Degond_etal_JCP11}.

This review is based on a sequence of papers about the derivation of hydrodynamic models for non momentum-conservative particle systems \cite{Degond_etal_MAA13, Degond_Motsch_M3AS08, Frouvelle_M3AS12, Motsch_Navoret_MMS11} and to a lesser extent, on  \cite{Barbaro_Degond_DCDSB13, Degond_etal_JNonlinearSci13, Degond_etal_note_submitted13, Degond_etal_preprint13, Degond_Liu_M3AS12, Degond_Motsch_JSP08, Degond_Motsch_JSP11, Degond_Yang_M3AS10, Frouvelle_Liu_SIMA12}. There is a vast literature on the mathematical modeling of collective motion and self-organization. We refer the reader, e.g. to \cite{Aoki_BullJapSocSciFish92, Baskaran_Marchetti_PRL10, Bertin_etal_JPA09, Chuang_etal_PhysicaD07, Couzin_etal_JTB02, Mogilner_etal_JMB03} and to the review \cite{Vicsek_Zafeiris_PhysRep12}. 

The paper is organized as follows. In section \ref{sec_particle}, we introduce the self-propelled particle dynamical system which will be at the heart of the present work. Then, in the limit of a large number of interacting particles, a mean-field model can be introduced and scaled in section \ref{sec_mf}. The hydrodynamic limit of the scaled mean-field model is studied in section \ref{sec_macro} and gives rise to the Self-Organized Hydrodynamic (SOH) model. Some properties of the SOH model are described in section \ref{sec_SOH_prop}. Local existence of smooth solution and characterization of weak solutions are investigated in section \ref{sec_SOH_existence}. A discussion of the model is given in section \ref{sec_discussion} and a conclusion is drawn in section \ref{sec_conclusion}.

\setcounter{equation}{0}
\section{Self-propelled particles interacting through alignment}
\label{sec_particle}

\begin{figure}[htbp]
\begin{center}
\input{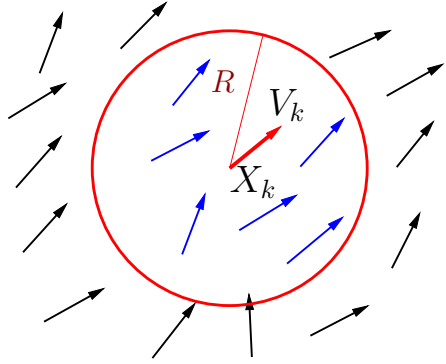}
\caption{The neighbors of the $k$-th particle located at $X_k$ with velocity $V_k$ (red arrow) are found in the ball enclosed by the red circle (ball centered at $X_k$ with radius $R$) and have velocities indicated by blue arrows. The average direction to which particle $k$ aligns (up to noise) is that of the sum of the blue vectors. }
\label{fig:Vicsek_ball}
\end{center}
\end{figure}

Our starting point is the Vicsek model \cite{Vicsek_etal_PRL95}. It describes self-propelled particles modelled as particles with constant velocity. They interact with each other by aligning with their neighbours up to a certain noise. Originally, the Vicsek model is a time discrete model, defined at discrete times separated by a constant time interval $\Delta t$.  The positions and velocities  of $N$ individuals at time $t^n = n \Delta t$ are denoted by $(X_k^n)_{k=1, \ldots, N}$ and $(\tilde V_k^n)_{k=1, \ldots, N}$. We denote by $m$ the spatial dimension, i.e. $X_k^n \in {\mathbb R}^m$ (in practice, $m=2$ or $3$). We assume that the particle speeds are constant and uniform, equal to $c>0$. Therefore, the velocity can be written $\tilde V_k^n = c V_k^n$, where $V_k^n$ lies on the sphere ${\mathbb S}^{m-1}$. The positions and velocities are evolved according to the following discrete algorithm: 
\begin{eqnarray}
& & \hspace{-1cm} 
X_k^{n+1} = X_k^n + c V_k^n \,  \Delta t , \label{eq:Vic_disc_X}\\
& & \hspace{-1cm} 
V_k^{n+1} = {\mathcal R}_{w_k^n,\theta_k^n} \bar V_k^n , \label{eq:Vic_disc_V}\\
& & \hspace{-1cm} 
\bar V_k^n = \frac{J_k^n}{|J_k^n|}, \quad  J_k^n = \sum_{ j, \, |X_j^n - X_k^n| \leq R } \,  V_j^n . \label{eq:Vic_disc_barV}
\end{eqnarray}
Eq. (\ref{eq:Vic_disc_X}) defines how the position of particle $k$ is updated from time $t^n$ to time $t^{n+1}$. It consists of a simple Euler discretization of the relation defining the velocity as the time-derivative of the position. Eq. (\ref{eq:Vic_disc_V}) states that the new particle velocity is set to the average neighbors' direction $\bar V_k^n$ up to a random noise, expressed by the operator ${\mathcal R}_{w_k^n,\theta_k^n}$. As expressed in (\ref{eq:Vic_disc_barV}), $\bar V_k^n$ is given by normalizing the vector obtained as the sum of the velocities $V_j^n$ of the particles lying in a ball or radius $R$ around the subject's position $X_k^n$ (see Fig. \ref{fig:Vicsek_ball}). The quantity $R$ is the interaction range of the subjects. For $\theta \in {\mathbb R}$ and $w \in {\mathbb S}^{m-2}_{\bar V}$ where ${\mathbb S}^{m-2}_{\bar V}$ is the $(m-2)$-dimensional unit sphere of the hyperplane orthogonal to $\bar V$, the operator ${\mathcal R}_{w,\theta} \bar V$ is defined by
\begin{eqnarray}
{\mathcal R}_{w,\theta} \bar V = \cos \theta \, \bar V + \sin \theta \, w. \label{eq:Vic_noise}
\end{eqnarray}
In (\ref{eq:Vic_disc_V}), $w_k^n$ are independent, uniformly distributed random vectors on the sphere ${\mathbb S}^{m-2}_{\bar V_k^n}$ and $\theta_k^n$ are independent, uniformly distributed random numbers in some interval $[0,D]$, with $0 \leq D \leq \pi$. 

In this model, the time step $\Delta t$ plays two different roles. The first one is that of a time discretization parameter (see Eq. (\ref{eq:Vic_disc_X})). The second one is that of an interaction frequency. Indeed, particles align with their neighbors at each time step. If $\Delta t$ is reduced by a factor $2$, the particles interact twice more frequently. Therefore, the limit $\Delta t \to 0$ in (\ref{eq:Vic_disc_X}), (\ref{eq:Vic_disc_barV})
does not lead to a well-posed problem because the number of interaction becomes infinite in a finite time interval. In order to bypass this problem, we decouple the two time scales and we introduce a collision frequency $\nu$ which is independent of the time step $\Delta t$. The original Vicsek model is recovered when $\nu = 1/\Delta t$. The modified discrete problem when  $\nu \not = 1/\Delta t$ is explicitly written in \cite{Degond_Motsch_M3AS08} and is omitted here. It can also be found in \cite{Czirok_etal_PRE96}. We directly introduce the time continuous Vicsek model, which is obtained as the limit $\Delta t \to 0$ of the modified time-discrete problem with collision frequency $\nu$. 

\begin{figure}[htbp]
\begin{center}
\input{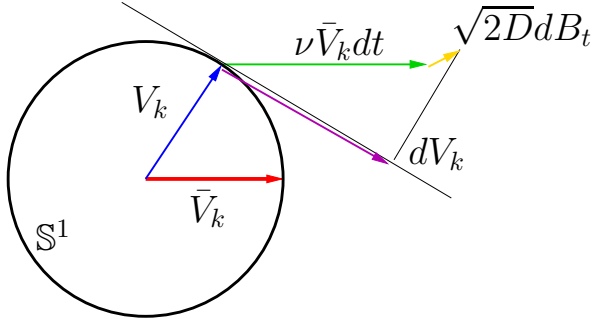}
\caption{Construction of the force term $dV_k(t)$ in dimension $m=2$: The velocity $V_k(t)$ (blue arrow) and the average neighbors' velocity $\bar V_k(t)$ (red arrow) are both vectors of the unit sphere ${\mathbb S}^1$ (black circle). To define the increment $dV_k(t)$, we add the interaction force vector $\nu \bar V_k(t) \, dt$ (green arrow) and a small random vector in ${\mathbb R}^2$ (yellow arrow) and project the resulting vector onto the line normal to $V_k(t)$ (purple arrow).}
\label{fig:cont_Vic}
\end{center}
\end{figure}

Let $X_k(t) \in {\mathbb R}^m$ and $V_k(t) \in {\mathbb S}^{m-1}$ be the position and velocity of the $k$-th particle at time $t$. The time-continuous version of the Vicsek model is written as follows. 
\begin{eqnarray} 
& & \hspace{-1cm} 
\dot X_k(t)   =  c \, V_k (t) , \label{eq:Vic_cont_X}\\
& & \hspace{-1cm} 
d V_k(t)   = P_{V_k^\bot} \circ (\nu \, \bar V_k dt + \sqrt{2D} \, dB^k_t) ,  \label{eq:Vic_cont_V} \\
& & \hspace{-1cm} 
\bar V_k = \frac{J_k}{|J_k|}, \quad J_k = \sum_{j,|X_j - X_k|\leq R} V_j 
, \label{eq:Vic_cont_barV}
\end{eqnarray}
where, for $V \in {\mathbb S}^{m-1}$, $P_{V^\bot} = \mbox{Id} - V \otimes V$ is the orthogonal projection onto the plane orthogonal to $V$. Eq. (\ref{eq:Vic_cont_X}) is the formal limit $\Delta t \to 0$ of (\ref{eq:Vic_disc_X}), while eq. (\ref{eq:Vic_cont_barV}) is the same as (\ref{eq:Vic_disc_barV}). The main change is Eq. (\ref{eq:Vic_cont_V}) for the evolution of $V_k$. It takes the form of a Stochastic Differential Equation (SDE). The projection operator $P_{V_k^\bot}$ ensures that the resulting solution $V_k(t)$ stays on the unit sphere, provided that the SDE is taken in the Stratonovitch sense (which is indicated by the symbol $\circ$). The first term inside the bracket is the interaction term. It corresponds to a force acting in the direction of $\bar V_k$ of intensity $\nu$. The second term is a white noise consisting of independent Brownians $B^k_t$ in ${\mathbb R}^2$ of intensity $\sqrt{2D}$. The fact that the projection $P_{V_k^\bot} \circ dB^k_t$ gives rise to a Brownian motion on the sphere provided that the SDE is taken in the Stratonovich sense can be found in e.g. \cite{Hsu_AMS02}. We can recover the original Vicsek model through a time discretization such that $\nu \Delta t = 1$ and the replacement of the white noise by the process described at (\ref{eq:Vic_noise}) \cite{Degond_Motsch_M3AS08}. The construction of the force term $dV_k(t)$ is illustrated in Fig. \ref{fig:cont_Vic}

Now, letting the number of particles $N \to \infty$, a mean-field model is obtained. This model is described in the following section.

\setcounter{equation}{0}
\section{Mean-field model and scaling}
\label{sec_mf}

The mean-field model describes the evolution of the one-particle distribution function $f(x,v,t)$, which depends on position $x \in {\mathbb R}^m$, velocity $v \in {\mathbb S}^{m-1}$ and time $t \geq 0$. The model is written as follows: 
\begin{eqnarray} 
& & \hspace{-1cm} 
\partial_t f + c v \cdot \nabla_x f = -   \nabla_v \cdot (F_f f) + D \Delta_v f , \label{eq:mfe_f}\\
& & \hspace{-1cm} 
F_f (x,t)  = \nu \, P_{v^\bot} \bar v_f(x,t), \quad \bar v_f(x,t) = \frac{{\mathcal J}_f(x,t)}{|{\mathcal J}_f(x,t)|} , \label{eq:mfe_F}\\
& & \hspace{-1cm}  
{\mathcal J}_f(x,t) = \int_{(y,w) \in {\mathbb R}^m \times {\mathbb S}^{m-1}} K\big( \frac{|y-x|}{R} \big) \,  f (y,w,t) \, w \,  dw \, dy  , \label{eq:mfe_j}
\end{eqnarray}
Eq. (\ref{eq:mfe_f}) is a Fokker-Planck equation. The left-hand side expresses the rate of change of $f$ due to the spatial transport of the particles with velocity $c v$ while the first term at the right-hand side takes care of the transport in velocity space due to the interaction force $F_f$. Finally, the last term at the right-hand side is a velocity diffusion term which arises as a consequence of the Brownian noise in particle velocities. Because $v$ lies on the sphere ${\mathbb S}^{m-1}$, $\Delta_v f$ stands for the Laplace-Beltrami operator on the sphere. The velocity diffusion coefficient $D$ is related to the Brownian noise intensity $\sqrt{2D}$ acting on the particles. The force term is proportional to the average neighbors' direction $\bar v_f(x,t)$ around $x$ at time $t$, projected on the hyperplane normal to $v$ through the action of $P_{v^\bot}$. The proportionality coefficient is nothing but the interaction frequency $\nu$. The projection $P_{v^\bot}$ ensures that the force term is normal to $v$, as it should to be consistent with the fact that $v\in {\mathbb S}^{m-1}$. The symbol \, $\nabla_v \cdot$ \, stands for the divergence of tangent vector fields to ${\mathbb S}^{m-1}$ (and later on \, $\nabla_v$ \ will denote gradients of scalar fields on ${\mathbb S}^{m-1}$; we recall that $\Delta_v = \nabla_v \cdot \nabla_v$). 

The local average neighbors' direction $\bar v_f(x,t)$ is obtained through the normalization of the vector ${\mathcal J}_f(x,t)$. This vector is computed by averaging the neighbors' velocities $w$ weighted by a function $K\big( \frac{|y-x|}{R}\big)$ depending on the distance between the particle and its neighbor $|x-y|$. The average is taken over the probability density $f(y,w,t)\, dy \, dw$. In the time-discrete case (section \ref{sec_particle}), we always took an abrupt cut-off of the interaction region, meaning that $K(\xi)$ is the indicator function of the interval $[0,1]$. However, smoother cut-offs represented by generic functions $K$ can be taken. The key assumptions on $K$ is that it should be positive, integrable, with a finite second moment. The quantity $R>0$ describes the typical interaction range. The derivation of the mean-field model (\ref{eq:mfe_f})- (\ref{eq:mfe_j}) from the discrete system (\ref{eq:Vic_cont_X})-(\ref{eq:Vic_cont_barV}) has been performed in a slightly different framework in \cite{Bolley_etal_AML11}. 
 
Macroscopic models are intended to describe the system at large time and space scales compared to those attached to the individuals and their interactions. In order to highlight the role of the time and space scales, we first non-dimensionalize the mean-field model (\ref{eq:mfe_f})- (\ref{eq:mfe_j}). We let $t_0$ be a time unit and $x_0 = c t_0$, $f_0 = 1/x_0^m$, $F_0 = 1/t_0$ be the associated space, distribution function and force units. We introduce the scaled variables $x'$; $t'$, etc. by the following relations:
$$ x = x_0 x', \quad t = t_0 t', \quad f = f_0 f', \quad F = F_0 F'. $$
Changing from variables $(x,v)$ to $(x',v')$ in system (\ref{eq:mfe_f})- (\ref{eq:mfe_j}) leads to the following model (dropping the primes for simplicity): 
\begin{eqnarray} 
& & \hspace{-1cm} 
\partial_t f + v \cdot \nabla_x f = -   \nabla_v \cdot (F_f f) + \bar D \Delta_v f , \label{eq:smfe_f}\\
& & \hspace{-1cm} 
F_f (x,t)  = \bar \nu \, P_{v^\bot} \bar v_f(x,t), \quad \bar v_f(x,t) = \frac{{\mathcal J}_f(x,t)}{|{\mathcal J}_f(x,t)|} , \label{eq:smfe_F}\\
& & \hspace{-1cm}  
{\mathcal J}_f(x,t) = \int_{(y,w) \in {\mathbb R}^m \times {\mathbb S}^{m-1}} K\big( \frac{|y-x|}{\eta} \big) \,  f (y,w,t) \, w \,  dw \, dy  , \label{eq:smfe_j}
\end{eqnarray}
where $\bar \nu = \nu t_0$, $\bar D = D t_0$, $\eta = R / x_0$.  

Now, we assume that at the chosen time scale $t_0$, we have $\bar \nu = {\mathcal O}\big( \frac{1}{\varepsilon} \big)$ and $\bar D = {\mathcal O}\big( \frac{1}{\varepsilon} \big)$, where $\varepsilon \ll 1$ is a small parameter.  Specifically, we let: 
\begin{equation}
\bar \nu = \frac{1}{\varepsilon}, \qquad \frac{\bar D}{\bar \nu} = d = {\mathcal O}(1).  
\label{eq:scaling}
\end{equation}
The parameter $\varepsilon$ measures the interaction time and interaction mean free path, i.e. the time and distance needed by a particle to make a finite change in direction of motion due to the interaction force. Equivalently, because the interaction force and noise operators are of the same order of magnitude thanks to (\ref{eq:scaling}), $\varepsilon$ is also the time needed by a particle to make a finite change of direction due to velocity diffusion. The time and space units $t_0$ and $x_0$ are macroscopic ones while the interaction time and mean free path are microscopic quantities. During a macroscopic time unit $t_0$, there are a  large (i.e. ${\mathcal O}(1/\varepsilon)$) number of interactions and  diffusions that sum up and contribute to making the corresponding operators large (specifically, both are ${\mathcal O}(1/\varepsilon)$ compared to the left-hand side of (\ref{eq:smfe_f})). At the macroscopic scale, the interaction force and diffusion terms must almost cancel each other to yield the order $1$ term at the left-hand side of (\ref{eq:smfe_f}). This cancellation is the mechanism which forms the 'Local Thermodynamical Equilibrium' detailed below in section \ref{sec_macro}. 

\begin{figure}[htbp]
\begin{center}
\input{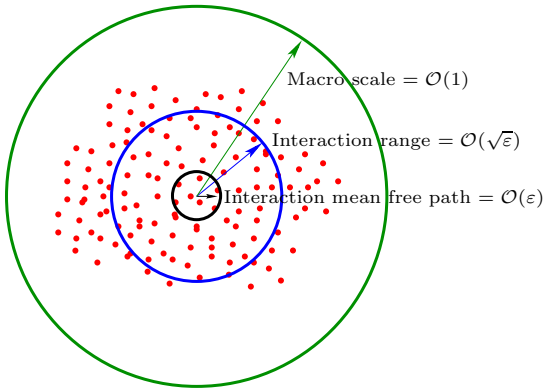}
\caption{The different scales of the problem in the case (ii) (see (\ref{eq:case_ii})): the microscopic scale is the interaction mean free path and is ${\mathcal O}(\varepsilon)$. The interaction range is the intermediate scale and is ${\mathcal O}(\sqrt \varepsilon)$. Finally the macroscopic scale is the scale of the whole system.}
\label{fig:scales}
\end{center}
\end{figure}

We also assume that $\eta \ll 1$ and we consider two scaling laws for $\eta$: 
\begin{eqnarray} 
& & \mbox{(i)} \quad \eta = {\mathcal O}(\varepsilon), \label{eq:case_i}\\
& & \mbox{(ii)} \quad \eta = \sqrt {\eta_0 \varepsilon}, \quad \mbox{with} \quad \eta_0>0, \quad \eta_0 = {\mathcal O}(1). \label{eq:case_ii}
\end{eqnarray}
In the first scaling (\ref{eq:case_i}), the interaction range is of the same order as the interaction mean free path and is therefore microscopic. With the second assumption (\ref{eq:case_ii}), the interaction range $\eta = {\mathcal O}(\sqrt {\varepsilon})$ is large compared to the interaction mean free path. It means that a particle interact with many more particles than just those that they are able to reach within an interaction time. In the macroscopic limit $\varepsilon \to 0$, the interaction range shrinks to $0$, meaning that it is smaller than a macroscopic quantity. Therefore, with the second scaling, the interaction range is intermediate between the microscopic scale and the macroscopic one. This choice of scales is illustrated in Fig. \ref{fig:scales}.

With these scaling assumptions, the scaled mean-field model is written: 
\begin{eqnarray} 
& & \hspace{-1cm} 
\varepsilon \big( \partial_t f^\varepsilon + v \cdot \nabla_x f^\varepsilon \big)  = -   \nabla_v \cdot (F^\eta_{f^\varepsilon} f^\varepsilon) + d \Delta_v f^\varepsilon , \label{eq:ssmfe_f}\\
& & \hspace{-1cm} 
F^\eta_f (x,t)  =  \, P_{v^\bot} \bar v_f^\eta(x,t), \quad \bar v_f^\eta(x,t) = \frac{{\mathcal J}_f^\eta(x,t)}{|{\mathcal J}_f^\eta(x,t)|} , \label{eq:ssmfe_F}\\
& & \hspace{-1cm}  
{\mathcal J}_f^\eta(x,t) = \int_{(y,w) \in {\mathbb R}^m \times {\mathbb S}^{m-1}} K\big( \frac{|y-x|}{\eta} \big) \,  f (y,w,t) \, w \,  dw \, dy  , \label{eq:ssmfe_j}
\end{eqnarray}
The goal is to study the limit $\varepsilon \to 0$. 
Since $\eta \to 0$ in both scalings (\ref{eq:case_i}) and (\ref{eq:case_ii}), we can expand $F^\eta_f (x,t)$ in powers of $\eta$ and get \cite{Degond_etal_MAA13}: 
\begin{eqnarray} 
& & \hspace{-1cm} 
F^\eta_f (x,t)  =  F^0_f (x,t) + \eta^2 F^1_f (x,t) + {\mathcal O}(\eta^4) , \label{eq:expanF_1} \\
& & \hspace{-1cm} 
F^0_f (x,t) = P_{v^\bot} u_f(x,t), \label{eq:expanF_2} \\
& & \hspace{-1cm} 
F^1_f (x,t) = \frac{k}{|j_f|} P_{v^\bot} P_{u_f^\bot} \Delta_x j_f , \label{eq:expanF_3} 
\end{eqnarray}
where the local density $\rho_f$, local current density $j_f$, and local average direction $u_f$ are defined by: 
\begin{eqnarray} 
& & \hspace{-1cm} 
\rho_f(x,t) = \int_{w \in {\mathbb S}^{m-1}}   f (y,w,t) \,  dw, \nonumber \\
& & \hspace{-1cm} 
j_f(x,t) = \int_{w \in {\mathbb S}^{m-1}}   f (y,w,t) \, w \,  dw  , \label{eq:rho_j} \\ 
& & \hspace{-1cm} 
u_f(x,t) = \frac{j_f(x,t)}{|j_f(x,t)|} . \label{eq:expanF_6} 
\end{eqnarray}
We have denoted by $k$ the second moment of $K$, i.e. 
$$ k = \frac{1}{2m} \int_{\xi \in {\mathbb R}^m} K(|\xi|) \, |\xi|^2 \, d \xi, $$
provided that $K$ is normalized to $1$ (without loss of generality), i.e. $\int_{\xi \in {\mathbb R}^m} K(|\xi|) \,  d \xi=1 $. 

Neglecting terms of order $\eta^4$, and defining the interaction (or collision) operator $Q(f)$ by:
\begin{eqnarray} 
& & \hspace{-1cm} 
Q(f) = - \nabla_v \cdot (P_{v^\bot} u_f f) + d \Delta_v f , \label{eq:def_Q}
\end{eqnarray}
we can write the system
\begin{eqnarray} 
& & \hspace{-1cm} 
\varepsilon \big( \partial_t f^\varepsilon + v \cdot \nabla_x f^\varepsilon \big)  +  \eta^2  \nabla_v \cdot (F^1_{f^\varepsilon} f^\varepsilon) = Q(f^\varepsilon) , \label{eq:scmfe_f}
\end{eqnarray}
Now, there are two cases according to which scaling assumption hypothesis (\ref{eq:case_i}) or (\ref{eq:case_ii}) is made. In case (\ref{eq:case_i}), the last term of the left-hand side of (\ref{eq:scmfe_f}) is of order $\varepsilon^2$ and can be neglected. In case (\ref{eq:case_ii}), we have $\eta^2 = \eta_0 \varepsilon$ and the last term of the left-hand side is of the same order as the other terms of the left-hand side. We can collect them and get the problem
\begin{eqnarray} 
& & \hspace{-1cm} 
\varepsilon \big( \partial_t f^\varepsilon + v \cdot \nabla_x f^\varepsilon   +  \eta_0  \nabla_v \cdot (F^1_{f^\varepsilon} f^\varepsilon) \big) = Q(f^\varepsilon) , \label{eq:sfmfe_f}
\end{eqnarray}
We note that we recover case (\ref{eq:case_i}) by just letting $\eta_0=0$ in (\ref{eq:sfmfe_f}). Therefore, case (\ref{eq:case_i}) becomes a sub-case of case (\ref{eq:case_ii}) and we now only consider (\ref{eq:sfmfe_f}) below. 
In the next section, we give the formal $\varepsilon \to 0$ limit of this model.

\setcounter{equation}{0}
\section{Self-Organized Hydrodynamics (SOH)}
\label{sec_macro}

\begin{figure}[htbp]
\begin{center}
\includegraphics[width=8cm]{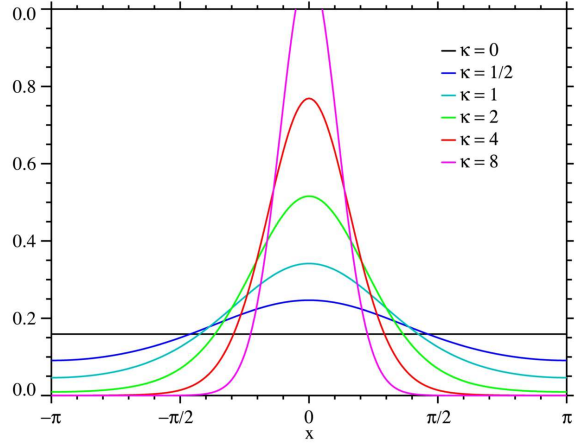}
\caption{The VMF distribution in dimension $m-1 = 1$, i.e. when the velocity variable is defined as $v = e^{i x}$ with $x \in ]- \pi, \pi ]$ (illustration taken from Wikipedia). The concentration parameter $\kappa = \frac{1}{d}$ controls the width of the distribution. For $\kappa = 0$, i.e. $d \to \infty$, the VMF distribution is very close to a uniform distribution (black horizontal line). When $\kappa$ increases or $d$ decreases, the height of the bump increases and its width decreases: the blue, blue-green, green, red and purple curves correspond to increasing values of $\kappa$ or decreasing values of $d$. The direction of the VMF corresponds to the center of the peak. Here, the direction of the VMF is $x=0$. }
\label{fig:VMF}
\end{center}
\end{figure}

The formal macroscopic limit $\varepsilon \to 0$ of (\ref{eq:sfmfe_f}) has been studied in \cite{Degond_etal_MAA13, Degond_Motsch_M3AS08}. We first state what are the equilibria, i.e. the solutions of $Q(f) = 0$. These solutions are important because $f^0 = \lim_{\varepsilon \to 0} f^\varepsilon$ (if it exists) is obviously an equilibrium thanks to (\ref{eq:sfmfe_f}). Since $Q$ only operates on the velocity variable, we first focus on the velocity dependence of these equilibria. Denote by $ {\mathcal E} = \{f \, | \,  Q(f) = 0 \}$ the set of equilibria. Then, it is possible to show that, under reasonable regularity assumptions,  $ {\mathcal E}$ is spanned by the so-called von Mises-Fisher (VMF) distributions, i.e. 
\begin{equation}
{\mathcal E} =  \{ \, v \to \rho M_{u}(v) \, \, \mbox{ for arbitrary } \, \, \rho \in {\mathbb R}_+, \, \,  u \in {\mathbb S}^{m-1} \}, 
\label{eq:equilibria}
\end{equation}
where 
$$ M_{u}(v) = Z_d^{-1} \, \exp \big( \frac{u \cdot v}{d} \big) , \quad \quad Z_d = \int_{v \in {\mathbb S}^{m-1}} \exp \big( \frac{u \cdot v}{d} \big)  \, dv. $$
We note that $M_{u}(v)$ is a probability density and that $Z_d$ does not depend on $u$. The element $u \in {\mathbb S}^{m-1}$ is called the direction of the VMF distribution while $\kappa = \frac{1}{d}$ is its concentration parameter. We also note that 
\begin{equation}
\int_{v \in {\mathbb S}^{m-1}} M_{u}(v) \, v  \, dv = c_1 \, u, \quad \quad c_1 = c_1(d) = \frac{\int_{v \in {\mathbb S}^{m-1}} \exp \big( \frac{u \cdot v}{d} \big)  (u \cdot v) \, dv}{\int_{v \in {\mathbb S}^{m-1}} \exp \big( \frac{u \cdot v}{d} \big)  \, dv}. 
\label{eq:current_equilibria}
\end{equation}
The quantity $c_1(d)$ does not depend on $u$ and satisfies $0 \leq c_1(d) \leq 1$. Eq. (\ref{eq:current_equilibria}) shows that the current associated to $M_u$ is directed and oriented along $u$ and its magnitude is defined by $c_1$. The function $\kappa = 1/d \in [0,\infty) \mapsto c_1(d) \in [0,1]$ is strictly increasing, onto. Small values of $c_1$ correspond to VMF distributions close to the uniform isotropic distribution, while values of $c_1$ close to $1$ correspond to VMF distributions close to Dirac deltas. Therefore, the parameter $c_1$ measures the degree of alignment of an ensemble of particles whose orientation $v$ is statistically defined by $M_u$. It is used as an order parameter in the study of phase transitions between disordered and aligned states. Fig.~\ref{fig:VMF} depicts a VMF distribution in dimension $1$. 

The VMF distribution allows us to write the collision operator $Q$ in the form of a Fokker-Planck operator. We have: 
$$ Q(f) (v) = d \, \nabla_v \cdot \left[ M_{u_f}(v) \nabla_v \left( \frac{f}{M_{u_f}}(v) \right) \right] . $$
We deduce a dissipation estimate
$$ \int_{v \in {\mathbb S}^{m-1}} Q(f) \, \frac{f}{M_{u_f}} \, dv = - d \int_{v \in {\mathbb S}^{m-1}} \left| \nabla_v \big( \frac{f}{M_{u_f}} \big) \right|^2 \, M_{u_f} \, dv \leq 0, $$
with an equality if and only if $f$ is an equilibrium i.e. $f \in {\mathcal E}$. The proof of (\ref{eq:equilibria}) relies on this inequality. Unfortunately, this dissipation estimate does not yield a H-theorem in the spatially inhomogeneous case because the multiplier $M_{u_f}$ depends on moments of $f$.  

From the determination of the equilibria, we know that $f^0 = \lim_{\varepsilon \to 0} f^\varepsilon$ (if it exists) is such that 
\begin{equation} 
f^0 (x,v,t) = \rho(x,t) M_{u(x,t)}(v), 
\label{eq:f0}
\end{equation}
where now $\rho = \rho(x,t)$ and $u=u(x,t)$ may have non-trivial dependences upon $(x,t)$. Indeed, since $Q$ operates only on the $v$ variable, it does not impose any constraint on the dependence of the parameters $\rho$ and $u$ of the equilibrium on $(x,t)$. 

In order to find this dependence, we need to introduce the second important concept, which is that of a collision invariant. A classical collision invariant is a function $\psi(v)$ such that 
$$ \int_{v \in {\mathbb S}^{m-1}} Q(f)(v) \, \psi(v) \, dv = 0 , $$
for all distribution functions $f$ with reasonable regularity (we will not dwell on this point here). The set of these collision invariants is denoted by ${\mathcal C}$ and is a vector space. The parameters  $(\rho,u) \in {\mathbb R}_+ \times {\mathbb S}^{m-1}$ of the equilibrium span a space of dimension $m$. To specify them, we need $m$ independent collision invariants, i.e. ${\mathcal C}$ must be of dimension $m$. Collision invariants are strongly related to conservation laws. For instance, in classical gas dynamics, the conservations of mass, momentum and energy provide $m+2$ independent collision invariants which allow to determine the $m+2$ independent parameters of the equilibrium Maxwellian distribution. In the present case, there exists only one conservation relation, namely the conservation of mass, which is associated to the collision invariant $\psi = 1$ (indeed, by Stokes formula, Eq. (\ref{eq:def_Q}) immediately leads to the fact that $ \int_{v \in {\mathbb S}^{m-1}} Q(f) \, dv = 0$). Due to the self-propelled character of the particles, there exists no momentum nor energy conservation nor any other kind of conservation. Consequently there exists no collision invariant independent from $\psi(v) = 1$. The dimension of ${\mathcal C}$ is $1$ and is strictly less that $m$. So, $m-1$ independent collision invariants are lacking to fully determine the parameters $\rho$ and $u$ of the VMF equilibria. 

To overcome this problem, we need to weaken the concept of a collision invariant. We define the concept of a 'Generalized Collision Invariant' (GCI) as follows. For any arbitrary vector $u \in {\mathbb S}^{m-1}$, we first define the linear operator ${\mathcal Q}_u$ by 
$$ {\mathcal Q}_u(f) (v) = d \, \nabla_v \cdot \left[ M_{u}(v) \nabla_v \left( \frac{f}{M_{u}}(v) \right) \right] . $$
we obviously have $Q(f) = {\mathcal Q}_{u_f}(f)$. Then, for any $u \in {\mathbb S}^{m-1}$, a function $\psi_u$ is a GCI associated to $u$ if and only if we have: 
\begin{equation}
 \int_{v \in {\mathbb S}^{m-1}} {\mathcal Q}_u(f)(v) \, \psi_u(v) \, dv = 0 , \quad \forall f \quad \mbox{such that} \quad P_{u^\bot} u_f = 0 . \label{eq:def_GCI}
\end{equation}
In other words, instead of testing $\psi$ against $Q(f)$ for all $f$, we restrict the set of test functions $f$ to those whose mean velocity direction $u_f$ is proportional to $u$ (and since both are unit vectors, this means $u_f = \pm u$) and use the linear operator ${\mathcal Q}_u$ instead of the nonlinear one $Q$. By taking a smaller set of test functions $f$, we are enlarging the set of possible solutions $\psi_u$. The concept of a GCI is now associated to a choice of a direction $u$. By a duality argument based on the fact that both the operator ${\mathcal Q}_u$ and the constraint $P_{u^\bot} u_f = 0$ (which is equivalent to $P_{u^\bot} j_f = 0$) are linear in $f$, it can be shown \cite{Degond_Motsch_M3AS08} that (\ref{eq:def_GCI}) is equivalent to the existence of a vector $\beta \in {\mathbb R}^m$ with $\beta \cdot u = 0$ such that 
\begin{equation} 
{\mathcal Q}_u^* (\psi_u) (v) = \beta \cdot v, 
\label{eq:GCI_adjoint}
\end{equation}
where ${\mathcal Q}_u^*$ is the $L^2$-adjoint of ${\mathcal Q}_u$. It is easy to see that the problem of finding $\psi_u$ is linear and consequently, the set ${\mathcal G}_u$ of the GCI associated to $u$ is a linear vector space. Since ${\mathcal Q}_u$ is a second order elliptic operator, Eq. (\ref{eq:GCI_adjoint}) is a second order elliptic problem posed on ${\mathbb S}^{m-1}$ whose solution can be found using Lax-Milgram theorem. This leads \cite{Degond_etal_JNonlinearSci13, Degond_etal_preprint13, Frouvelle_M3AS12} to the fact that ${\mathcal G}_u$ is an $m$-dimensional vector space given by
$${\mathcal G}_u = \{ v \mapsto C + h(u \cdot v) \, \beta \cdot v, \mbox{ with arbitrary }  C \in {\mathbb R} \mbox{ and } \beta \in {\mathbb R}^m \mbox{ with } \beta \cdot u = 0 \}. $$
The unique scalar function $h$ is defined as follows. Denote by $u \cdot v = \cos \theta$. Then, $h(\cos \theta) = g(\theta) / \sin \theta$ where $g(\theta)$ is the unique solution in the space $V$ of the elliptic problem $\tilde L^* g = \sin \theta$, with 
$$ \tilde L^*g(\theta)=-\sin^{2-m}\theta \, \,  e^{-  \frac{\cos\theta}{d}} \, \,  \frac{d}{d \theta} \big(\sin^{m-2}\theta \, \,  e^{ \frac{\cos\theta}{d}} \, \,\frac{dg}{d \theta}(\theta) \big)+\frac{m-2}{\sin^2\theta} \, g(\theta), $$
and 
$$ V = \{ g \, \, \, | \, \, \,(m-2)\, (\sin\theta)^{\frac{m}{2}-2} \, g \in L^2(0,\pi), \, \, \, (\sin\theta)^{\frac{m}{2}-1} \,  g \in H^1_0(0,\pi) \}, $$
(denoting by $H^1_0(0,\pi)$ the Sobolev space of functions which are square integrable as well as their derivative and which vanish at the boundary). Since a function of ${\mathcal G}_u$ is determined by the scalar $C$ and the vector $\beta$ such that $\beta \cdot u = 0 $, it depends on $m$ independent parameters and forms a linear space of dimension $m$. 

Now, we can perform the formal limit $\varepsilon \to 0$, assuming that $f^\varepsilon \to f^0$ as smoothly as needed (i.e. with a regularity such that all the formal computations below can be rigorously justified). We already know that $f^0$ is an equilibrium  given by (\ref{eq:f0}). Now, integrating (\ref{eq:sfmfe_f}) with respect to $v$, using Stokes' formula on the sphere and taking the limit $\varepsilon \to 0$, we find the continuity equation
\begin{eqnarray*} 
& & \partial_t \rho + \nabla_x \cdot (c_1 \rho u)  = 0  .  
\end{eqnarray*}
The fact that the particle flux at equilibrium is expressed by $c_1 \rho u$ comes from (\ref{eq:current_equilibria}). 

In order to find the equation satisfied by $u$, we multiply (\ref{eq:sfmfe_f}) by the vector-valued GCI $\Psi_{u_{f^\varepsilon}}$, where, for an arbitrary vector $u \in {\mathbb S}^{m-1}$, $\Psi_u(v) = h(u \cdot v) P_{u^\bot} v$, and take the limit $\varepsilon \to 0$. We note that, as a consequence of the definition of the GCI, we have $\int_{v \in {\mathbb S}^{m-1}} Q(f^\varepsilon) \, \Psi_{u_{f^\varepsilon}} \, dv = 0$. Consequently, the right-hand side of (\ref{eq:sfmfe_f}), which is singular as $\varepsilon \to 0$, disappears after integration against $\Psi_{u_{f^\varepsilon}}$. After some algebra, the left-hand side gives rise to the following equation for $u$: 
\begin{eqnarray*} 
& &  \rho \, \left( \partial_t u + c_2 (u \cdot \nabla_x) u \right) + d  \,  P_{u^\bot} \nabla_x \rho = c_3 P_{u^\bot} \Delta_x (\rho u) 
\end{eqnarray*}
with 
\begin{eqnarray*} 
c_2=  \frac{\int_0^\pi \cos \theta\, h(\cos\theta) \, e^{\frac{\cos\theta}{d}} \, 
\sin^m \theta \, d\theta}{\int_0^\pi h(\cos\theta) \,e^{\frac{\cos\theta}{d}} \, \sin^m \theta \, d\theta}\, , \quad \quad c_3 = \eta_0 k \big( (m-1) d + c_2 \big). 
\end{eqnarray*}

We summarize the result in the following:

\begin{theorem}
As $\varepsilon \to 0$, we have $f^\varepsilon \to f^0$ formally, where $f^0$ is an equilibrium (\ref{eq:f0}) whose parameters $\rho(x,t)$ and $u(x,t)$ satisfy the following 'Self-Organized Hydrodynamics' (SOH) system: 
\begin{eqnarray} 
& & \partial_t \rho + \nabla_x \cdot (c_1\rho u)  = 0, \label{eq:SOH_rho}   \\
& &  \rho \, \left( \partial_t u + c_2 (u \cdot \nabla_x) u \right) + d  \,  P_{u^\bot} \nabla_x \rho = c_3 P_{u^\bot} \Delta_x (\rho u)  , \label{eq:SOH_u}\\
& & |u| = 1 . \label{eq:SOH_norm}
\end{eqnarray}
\label{thm:SOH}
\end{theorem}

In the next section, we review some properties of the SOH model, we state the available existence results and we address the question of its numerical approximation.

\setcounter{equation}{0}
\section{Properties of the SOH model}
\label{sec_SOH_prop}

The SOH system (\ref{eq:SOH_rho})-(\ref{eq:SOH_norm}) bears strong similarities with the isothermal compressible Navier-Stokes (NS) model, as recalled below: 
\begin{eqnarray} 
& & \partial_t \rho + \nabla_x \cdot (\rho u)  = 0, \label{eq:CNS_rho}   \\
& &  \rho \, \left( \partial_t u + (u \cdot \nabla_x) u \right) + T  \, \nabla_x \rho = \nu \, \nabla_x \cdot (\nabla_x u + \nabla_x u^*)  , \label{eq:CNS_u}
\end{eqnarray}
where $\rho(x,t) \geq 0$ and $u(x,t) \in {\mathbb R}^m$ are the gas density and mean velocity, $T$ is the temperature (the analog of our $d$) and $\nu$ is the viscosity (the analog of our $c_3$). The quantity $\nabla_x u$  is the tensor gradient of the vector field $u$ and the exponent '$*$' denotes the transpose of a tensor. We adopt the notation \, $\nabla_x \cdot$ \, indifferently for divergence of vectors and tensors. We have written the momentum equation (\ref{eq:CNS_u}) in non-conservative form to highlight the analogy with (\ref{eq:SOH_u}) but it is more natural to express it in conservative form. This form is obtained by multiplying (\ref{eq:CNS_rho}) by $u$ and adding to (\ref{eq:CNS_u}). This leads to 
\begin{eqnarray} 
& &  \hspace{-1cm}
\partial_t (\rho u) + \nabla_x \cdot (\rho u \otimes u) + \, \nabla_x (\rho T) = \nabla_x \cdot \big( \nu ( \, \nabla_x u + \nabla_x u^* \, )\big)  , \label{eq:CNS_u_cons}
\end{eqnarray}
where $u \otimes u$ stands for the tensor product of the vector $u$ by itself. In the inviscid case $\nu=0$, the conservative form allows to define solutions with jumps thanks to the Rankine-Hugoniot condition. Additionally, the entropy condition, which we will not recall here, permits the selection of a single solution among the possible weak solutions.

Both the SOH and NS models are constructed as nonlinear systems of first order equations perturbed by diffusion. The nonlinear first order parts consist of the left-hand sides of (\ref{eq:SOH_rho}), (\ref{eq:SOH_u}) on the one-hand and (\ref{eq:CNS_rho}), (\ref{eq:CNS_u}) on the other hand. The diffusion operators appear in the velocity equations (the right-hand sides of (\ref{eq:SOH_u}) and (\ref{eq:CNS_u}) respectively). Additionally, both nonlinear first order parts are hyperbolic. This is standard for the NS model: its first order part is the isothermal compressible Euler equations whose characteristic velocities are $\lambda_{\pm} = (u \cdot \xi) \pm \sqrt T$ and $\lambda_0 = u \cdot \xi$ where $\xi$ is the unit vector in the direction of propagation of the characteristic wave. In the case of the SOH model, these characteristic velocities have been computed in \cite{Degond_Motsch_M3AS08} and are given by 
\begin{equation} 
\lambda_\pm = \frac{1}{2} \left[ (c_1 + c_2) \, u \cdot \xi  \pm \left( (c_2 - c_1)^2 \, (u \cdot \xi)^2 + 4d \, (1-(u \cdot \xi)^2) \right)^{1/2} \right], \quad \lambda_0 = c_1 \, u \cdot \xi . 
\label{eq:char_vel}
\end{equation}

The first important difference between the two models is that the SOH model involves the geometric constraint (\ref{eq:SOH_norm}) which imposes the velocity $u$ to be of unit norm. By contrast, the standard NS model does not involve such a constraint. We note that, for smooth solutions of the SOH model, this geometric constraint is satisfied at all times provided it is satisfied initially, i.e. if $|\, u_{|t=0} \, | =1$. Indeed, $P_{u^\bot}$ multiplies the last term of the left-hand side and the right-hand side of (\ref{eq:SOH_u}). Therefore,  multiplying scalarly (\ref{eq:SOH_u}) by $u$, we get:
$$ \big(\partial_t + c_2 (u \cdot \nabla_x) \big)\big( \frac{|u|^2}{2} \big) = \big[\big(\partial_t + c_2 (u \cdot \nabla_x) \big) u\big] \cdot u = 0 , $$
which shows that $|u|^2$ is a conserved quantity along the characteristics of the vector field $c_2 u(x,t)$. If  $|u|^2$ is initially equal to $1$ uniformly, it stays equal to $1$ uniformly at future times. This computation requires the solution to be smooth. The preservation of the norm along discontinuous trajectories is still a conjecture  at this time.  

The geometric structure is brought to the model by the multiplication of the last term of the left-hand side and of the right-hand side of (\ref{eq:SOH_u}) by the projection operator $P_{u^\bot}$. The matrix $P_{u^\bot} = \mbox{Id} - u \otimes u$ being a non-trivial function of $u$, this multiplication introduces non-conservative products. Indeed, the corresponding terms are spatial derivatives of non-trivial functions of the unknowns $\rho$ and $u$, multiplied by non-trivial functions of these unknowns. Non-conservative hyperbolic systems are reviewed in \cite{LeFloch_CPDE88}. They have some unpleasant features such as the impossibility of defining shock relations for discontinuous solutions by means of the Rankine-Hugoniot condition. As a consequence, the shock speeds are not well-defined and there is no simple criterion to single out a particular solution among the possible ones. 

In the SOH model, the fluid mean velocity is not given by $u$ but rather, by $c_1 u$. The vector $u$ gives the average orientation of the particles. Most continuum models of self-propelled particles available in the literature, such as \cite{Toner_Tu_PRL95, Toner_etal_AnnPhys05, Tu_etal_PRL98}, use the mean velocity as a variable of the model, instead of the mean orientation. As a consequence, in these models, $u$ is not constrained to be of unit norm, and they bear a closer resemblance with the NS than the SOH model presented here. However, these models are constructed on phenomenological bases, while the SOH model is derived through a rigorous hydrodynamic limit of the underlying mean-field model. 

The second important difference between the SOH model and the standard NS model is the constant $c_2$ which, in general, is different from $c_1$. In the present case, we can prove \cite{Degond_Motsch_M3AS08} that $0 < c_2 \leq c_1$ (where $c_2 = c_1$ if and only if the noise intensity $d=0$). Since $c_1 u$ is the flow velocity and $c_2 u$ is the velocity at which the velocity itself is transported, this means that velocity is transported upstream the flow. The transport of velocity is a transport of the information about how agents should adjust their velocity to accommodate for the presence of other agents in the front. This is a situation similar to vehicular traffic (see e.g. \cite{Aw_Rascle_SIAP00}), where information about how drivers should adapt their velocity to the vehicles in the front propagates upstream the flow and even, in congestion situations, opposite to the flow direction. 

When anisotropic vision is considered, like in \cite{Frouvelle_M3AS12}, arbitrary values of $c_2 \in {\mathbb R}$ can be generated. Anisotropic vision means that the interaction kernel $K$ does not only depend on the distance $|y-x|$ between the subject located at $x$ and its partner located an $y$ but also on the bearing angle, i.e. the angle $\widehat{(v, y-x)}$ between the subject's velocity $v$ and the line of sight $y-x$ under which he sees his partner. In particular, forward vision corresponds to situations where $K$ is lower for obtuse bearing angles, and backwards vision, where $K$ is larger. In the isotropic vision case discussed so far, $K$ only depended on the distance $|y-x|$. In \cite{Frouvelle_M3AS12}, it is shown that forward vision results in general in lower values of $c_2$ than isotropic vision, while backwards vision results in larger values. In the case of forward vision, for large enough values of the noise intensity $d$, one can even have $c_2 <0$, meaning that velocity is transported in opposite direction to the flow. As noted above, this is similar to car traffic in congested situations. In the case of backwards vision, for large values of $d$, we can have $c_2 > c_1$, meaning that information propagates downstream the flow. It is shown in \cite{Degond_Hua_JCP13} that $c_2 < c_1$ increases the ability of the model to generate density concentrations and shocks, while this tendency is lowered in the case $c_2 > c_1$. As pointed out above, the case $c_2 < c_1$ bears analogies with vehicular traffic. By contrast, the case $c_2 > c_1$ is exemplified in locusts' coordinated mass migration. The locust species {\em Schistocerca gregaria} is a cannibalistic species and individuals try to avoid being bitten by others in their back \cite{Bazazi_etal_CurrBiol08}. This is an example where backwards vision (in this case, the sensory mechanism is not vision but rather abdomen innervation) controls collective motion. 

The SOH model is not Galilean-invariant because of the constraint $|u|=1$. Indeed, it is not possible to translate the velocity $u$ by a uniform velocity $V$ and keep this constraint. That $c_2 \not = c_1$ also induces some non-Galilean effects. Indeed, if we perform the same manipulations as those leading to the conservative form (\ref{eq:CNS_u_cons}) of the NS model, we are led to: 
\begin{eqnarray} 
& &  \hspace{-1cm}
\partial_t (\rho u) + \nabla_x \cdot (c_1 \rho u \otimes u) + (c_2 - c_1) \, ( u \cdot \nabla_x) u  + P_{u^\bot} \nabla_x (d \rho) = c_3 P_{u^\bot} \Delta_x (\rho u)  , 
\label{eq:SOH_u_cons}
\end{eqnarray}
The extra term $(c_2 - c_1) \, ( u \cdot \nabla_x) u$ can be understood as a velocity-dependent extra pressure term. The fact that pressure is velocity dependent is a signature of a non-Galilean model, because in Galilean-invariant fluids, the pressure is an intrinsic property of the fluid, which is the same in all reference frames. Another manifestation of the non Galilean-invariant character of the model is the expression of the sound speed $c_s$, which corresponds to the square root term in the expression (\ref{eq:char_vel}) of the characteristic speeds $\lambda_\pm$, i.e.
\begin{equation}
c_s = \left( (c_2 - c_1)^2 \, (u \cdot \xi)^2 + 4d \, (1-(u \cdot \xi)^2) \right)^{1/2}. 
\label{eq:sound_speed}
\end{equation}
 This expression depends on the angle between the propagation direction $\xi$ and the velocity $u$. In Galilean-invariant fluids, the sound speed is independent of the propagation direction. Like our model, the previously proposed  continuum models of swarming are not Galilean-invariant and the effect of non Galilean-invariance on sound propagation is reviewed in \cite{Tu_etal_PRL98}. 

\begin{figure}[htbp]
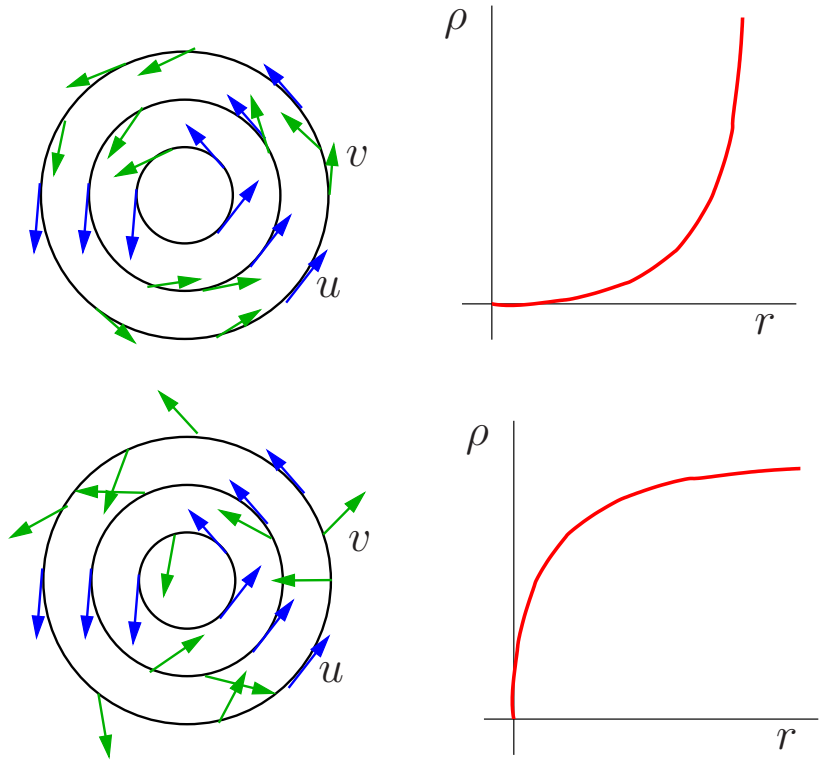

\begin{center}
\input{Mill_1_al.pstex_t} \hspace{1cm}
\input{Mill_2_al.pstex_t}

\vspace{0.5cm}
\input{Mill_1_rn.pstex_t} \hspace{1cm}
\input{Mill_2_rn.pstex_t}
\caption{Schematics of a milling solution: left: mean orientation $u$ (blue arrows) and individual particle velocities $V_k^n$ (green arrows) ; right: density profile $\rho$ as a function of the distance to the origin $r = |x|$. Top figure: small noise case $c_2/d > 1$: the density is convex and the particle velocities $V_k^n$ follow closely the mean orientation $u$. Bottom figure: large noise case $c_2/d < 1$: the density is concave and the particle velocities $V_k^n$ deviate strongly from the mean orientation $u$. }
\label{fig:mills}
\end{center}
\end{figure}

Finally, we note that special stationary solutions are given by mills, i.e. solutions of the form
$$\rho(x) = \rho(|x|) =  \rho_0 \, ( r \, / \, r_0 )^{c_2/d} \, , \quad u = x^\bot/|x|, $$
where $\rho_0>0$ and $r_0>0$ are arbitrary constants. In such solutions, the average velocity corresponds to a spinning about the origin at a constant speed, while the density profile is unbounded and increases like a power law of the distance to the origin. According to the position of $c_2/d$ with respect to $1$, the density profile is convex or concave. Convex density profiles are associated to small noise and consequently, to a clear spinning organization of the individual particle velocities. By contrast, at large noise, the density profile becomes concave. Due to the large noise, the individual velocities show large deviations around the average spinning velocity, and the organization of the individual particles in a milling pattern is less clear. These are illustrated in Fig. \ref{fig:mills}. The question of the stability of these milling solutions is an open problem. Milling is a common yet intriguing social behavior in fish populations \cite{Domeier_Colin_BullMarSci97}. 

\setcounter{equation}{0}
\section{Local existence of smooth solutions and selection criterion for discontinuous solutions}
\label{sec_SOH_existence}

In this section we state the existence results of \cite{Degond_etal_MAA13}. We first start with the case of dimension $m=2$. In this case, we can take $c_3 \geq 0$. We assume that the spatial domain is the two dimensional torus, i.e. $x \in \Pi^2$, with $\Pi = [0,1]$ with periodic boundary conditions. The initial data $(\rho_0, u_0)$ are such that $\rho_0 >0$ and $|u_0| = 1$. We write $u = (\cos \varphi, \sin \varphi)$. We have the 

\begin{theorem} \cite{Degond_etal_MAA13}
We assume that the initial data belong to $H^s(\Pi^2)$ with $s > 2$. Then, there exists a time $T > 0$ and a unique solution $(\rho, \varphi) \in L^\infty([0,T],H^s(\Pi^2)) \cap H^1([0,T],$ $H^{s-1}(\Pi^2))$ of the SOH model such that $\rho$ remains positive. If, in addition, $c_3 > 0$, then, the solution also belongs to $L^2([0,T], H^{s+1}(\Pi^2))$.
\label{thm:exist2D}
\end{theorem}

The proof relies on the fact that the first order part of the SOH model admits a symmetrizer. This symmetrizer allows the development of energy estimates.  

We now turn to the dimension $m=3$. In this case, we use global spherical coordinates and write  $u = (\sin \theta \, \cos \varphi, \sin \theta \, \sin \varphi, \cos \theta)$. Due to the singularity of this coordinate system for $\theta = 0$ and $\theta = \pi$, we choose initial velocities such that $(\theta_0, \varphi_0) \in [\theta_m, \theta_M] \times [0,2 \pi]$, with $0 < \theta_m < \theta_M < \pi$. In the case $c_3 = 0$, we can use the finite speed of propagation of hyperbolic models to guarantee that the solution $\theta$ will stay away from the singular points $\theta = 0$ and $\theta = \pi$ during some interval of time. Because diffusion involves propagation at infinite speed, this property is lost in the case $c_3 > 0$ and the proof breaks down. This is why we restrict ourselves to the inviscid case $c_3 = 0$ in dimension $m=3$. Most likely, this restriction could be waived at the expense of technicalities, such as changing the local coordinates on the sphere near the singular points but these developments are left to future work.  

Again, a periodic domain $\Pi^3$ is chosen as spatial domain and the initial density is supposed positive $\rho_0 >0$. 
Then, in the case $m=3$ and $c_3 = 0$, the theorem reads as follows 

\begin{theorem} \cite{Degond_etal_MAA13} 
We assume that the initial data $(\rho_0, \theta_0, \varphi_0)$ belong to $H^s(\Pi^3)$ with $s > 5/2$ and that $\rho_0 >0$, $\sin \theta_0 >0$. Then, there exists a time $T > 0$ and a unique solution $(\rho, \theta, \varphi) \in L^\infty([0,T],H^s(\Pi^3)) \cap H^1([0,T],H^{s-1}(\Pi^3))$ of the SOH model such that $\rho$ remains positive.
\label{thm:exist3D}
\end{theorem}

We show that the inviscid SOH model in dimension $m=3$ can be written in the form of symmetrizable hyperbolic system. Existence and uniqueness follows from the classical theory of symmetrizable hyperbolic systems. 

These local existence results for smooth solutions do not give any information about the existence and uniqueness of non-smooth solutions. As a non-conservative model, the SOH may have multiple shock velocities \cite{LeFloch_CPDE88} and there is no obvious analytic criterion to single out one particular solution. Following \cite{LeFloch_CPDE88}, given two states $(\rho_\ell, u_\ell)$ and $(\rho_r, u_r)$ separated by a spatial discontinuity, there are as many shock relations as possible paths in the state space spanned by $(\rho, u)$ (i.e. ${\mathbb R}_+ \times {\mathbb S}^{m-1}$) connecting these two states. Most non-conservative systems, like the SOH model, are obtained by some coarse graining procedure from an underlying 'microscopic' model. What is the correct path is an information which has been lost in the coarse graining procedure. In principle, this information could be retrieved if one could 'interrogate' the microscopic model. Unfortunately, this is not doable in practice. 

Here, the coarse graining procedure is the hydrodynamic limit and the microscopic model is the time-continuous particle system (\ref{eq:Vic_cont_X}), (\ref{eq:Vic_cont_barV}). Therefore, we can compare how the SOH and particle systems resolve a Riemann problem and try to find experimentally what are the correct shock relations for the SOH model. In fact the problem is considerably simplified by the remark made in \cite{Motsch_Navoret_MMS11}. It is based on the observation that the SOH model can be formally obtained as a relaxation limit of a conservative model, the so-called Relaxed SOH model (RSOH). The RSOH model is written as follows: 
\begin{eqnarray} 
& & \partial_t \rho^\delta + \nabla_x \cdot (c_1\rho^\delta u^\delta)  = 0, \label{eq:RSOH_rho}   \\
& &  \partial_t (\rho^\delta u^\delta) + \nabla_x \cdot (c_2 \, \rho^\delta u^\delta \otimes u^\delta )  + d  \,  \nabla_x \rho^\delta =   - \frac{1}{\delta} \, \rho^\delta \, (1-|u^\delta|^2) \, u^\delta , \label{eq:RSOH_u}
\end{eqnarray}
where $\delta \ll 1$ is a relaxation parameter. Here we have considered the inviscid case $c_3 = 0$ for simplicity. It is possible to formally show \cite{Motsch_Navoret_MMS11} that $(\rho^\delta, u^\delta) \to (\rho, u)$ as $\delta \to 0$  where $(\rho, u)$ is a solution of the SOH model. The RSOH model is hyperbolic if and only if $c_2 \geq c_1$. By contrast, in the case where $c_2 < c_1$, it is only conditionally hyperbolic, when 
$$|u| \leq \sqrt{\frac{d}{\frac{c_2}{c_1} ( 1 - \frac{c_2}{c_1})}}. $$
The SOH model being a relaxation limit of the larger conservative system (\ref{eq:RSOH_rho}), (\ref{eq:RSOH_u}), one may think that it is amenable to the relaxation theory for systems of conservation laws \cite{Chen_etal_CPAM94}. In fact, it is not the case. Indeed, in \cite{Chen_etal_CPAM94}, a key hypothesis is that the relaxed system is also a system of conservation laws. Here the system is non-conservative and the theory does not apply. In particular, the ${\mathcal O}(\delta)$ correction terms to the SOH model that occur in a Chapman-Enskog expansion of the RSOH system can be computed. They are not diffusive terms by contrast to those which are obtained in the theory of \cite{Chen_etal_CPAM94}. Indeed, the geometric constraint in the SOH model opens a wealth of novel features, most of them being still unexplored. 

From the RSOH model, a relaxation scheme for the SOH model can be designed. It consists of a time splitting method. Given an approximate solution $(\rho^n, u^n)$ of the SOH model at time $t^n = n \Delta t$, the first step of the splitting consists in solving the non relaxation part of the RSOH model (\ref{eq:RSOH_rho}), (\ref{eq:RSOH_u}), i.e. the system 
\begin{eqnarray} 
& & \partial_t \rho + \nabla_x \cdot (c_1\rho u)  = 0,  \label{eq:NRSOH_rho} \\
& &  \partial_t (\rho u) + \nabla_x \cdot (c_2 \, \rho u \otimes u )  + d  \,  \nabla_x \rho =   0 , \label{eq:NRSOH_u}
\end{eqnarray}
over one time step $\Delta t$ with initial condition $(\rho^n, u^n)$ by a standard shock-capturing scheme such as the Rusanov method \cite{Rusanov_JCMP61}. This leads to intermediate values $(\tilde \rho^{n+1}, \tilde u^{n+1})$. Then, in the second step of  the splitting, the relaxation part of the RSOH model is solved over one time-step $\Delta t$ with initial condition $(\tilde \rho^{n+1}, \tilde u^{n+1})$. This relaxation part reads: 
\begin{eqnarray*} 
& & \partial_t \rho^\delta   = 0,   \\
& &  \partial_t (\rho^\delta u^\delta) =   - \frac{1}{\delta} \, \rho^\delta \, (1-|u^\delta|^2) \, u^\delta . 
\end{eqnarray*}
The solution of this system can be explicitly computed and in the limit $\delta \to 0$, just reduces to the normalization of the velocity issued from the first step of the splitting. This leads to the new value $(\rho^{n+1},u^{n+1})$ of the solution at time $t^{n+1}$: 
\begin{eqnarray} 
\rho^{n+1} = \tilde \rho^{n+1}, \quad \quad u^{n+1} =\frac{\tilde u^{n+1}}{|\tilde u^{n+1}|}. 
\label{eq:normalization}
\end{eqnarray}

In the situation where the RSOH model is not hyperbolic, the method can still be used. Indeed, in the splitting method, the solution of the non-relaxation part of the RSOH model (\ref{eq:NRSOH_rho}), (\ref{eq:NRSOH_u}) is immediately reprojected onto a solution of the SOH model through the normalization (\ref{eq:normalization}). So, an instability does not have time to develop before the solution is reprojected into a solution of the SOH hyperbolic model. The stability of this solution methodology has been experimentally demonstrated in \cite{Motsch_Navoret_MMS11} but a rigorous proof of this property is still lacking. Now, a pending question still remains: how to practically solve the non-relaxation part of the RSOH model (\ref{eq:NRSOH_rho}), (\ref{eq:NRSOH_u}) in a non-hyperbolic situation~? The answer is easy in the context of e.g. the Rusanov method. Indeed, in this method, the computation of the characteristic speeds is only needed in the determination of the numerical viscosity. Using the 'wrong' numerical viscosity is not detrimental as long as it does not become too small. On the other hand, overestimating the numerical viscosity may deteriorate the quality of the solution, but does not threaten the stability of the method. In the model, the breakdown of hyperbolicity occurs by the appearance of a negative real number inside the square root in the expression of the sound speed (\ref{eq:sound_speed}). Taking the absolute value of this negative number allows to assign a real value to the square root and consequently, to set a value to the numerical viscosity. Using this value apparently provides correct solutions without the appearance of any instability problem~\cite{Degond_Hua_JCP13}. However, a rigorous basis to this methodology is still unavailable.  

It has been experimentally discovered in \cite{Motsch_Navoret_MMS11} that using this relaxation scheme to solve the Riemann problem for the SOH model provides extremely good approximations of the corresponding solution of the particle system. By contrast, any standard shock capturing method adapted to non-conservative models provides wrong solutions. Therefore, although we are not able to provide an analytic criterion for the selection of the 'correct' discontinuous solutions, an experimental way to compute them numerically exists. Details and numerical evidence can be found in \cite{Motsch_Navoret_MMS11}. 

To conclude this section, we present some numerical simulations in Fig. \ref{fig:simules}. This simulations show that the SOH model provides a fairly good approximation to the solution of the particle model even for long simulation times. 

\pagebreak

\begin{figure}[htbp]
\begin{center}
\includegraphics[height=5cm]{simu_t0.eps}

\vspace{0.5cm}
\includegraphics[height=5cm]{simu_t30.eps}

\vspace{0.5cm}
\includegraphics[height=5cm]{simu_t60.eps}
\caption{Comparison between the particle model (\ref{eq:Vic_cont_X}), (\ref{eq:Vic_cont_barV}) (left panels) and the SOH model (right panels). Top panel: time $t=0$. Middle panel: time $t=30$. Bottom panel: time $t=60$. The speed of the particles is $1$ and the size of the domain is $20$. The average density in each cell is represented by a color code from low density (light yellow) to large density (dark red). The average velocity in each cell is represented by an arrow in the center of the cell. For the SOH model, the magnitude of the average velocity is constant equal to $c_1$. We observe that the agreement is fairly good in spite of a noisy initial condition and a large simulation time.}
\label{fig:simules}
\end{center}
\end{figure}

\setcounter{equation}{0}
\section{Discussion}
\label{sec_discussion}

We have presented the derivation of the SOH model in the simplest situation where the collision frequency and noise are constant. Several generalizations have been proposed. In \cite{Degond_Motsch_M3AS08}, the collision frequency $\nu$ depends on the angle $u \cdot \bar v_f$ between the particle velocity and the neighbor's average orientation. All the results given above extend to this case, except for the value of the constants $c_1$, $c_2$ and for the coefficient multiplying the pressure term $P_{u^\bot} \nabla_x \rho$. We denote by $\Theta$ this coefficient (in the case discussed here we have $\Theta = d$).
In \cite{Degond_Motsch_M3AS08}, the derivation of the SOH model was made in dimension $3$ only. In \cite{Frouvelle_M3AS12}, an arbitrary dimension $m \geq 2$ is considered and both the collision frequency $\nu$ and the noise $d$ are made dependent of the local density $\rho_f$. Again, all the present results extend to this case, except that now, $c_1$, $c_2$ and $\Theta$ become functions of the density. Expressions of these constants are given in arbitrary dimensions and asymptotic developments of them when $d \to 0$ and $d \to \infty$ are provided. Additionally, it is shown that, with density dependent coefficients, the SOH model may lose its hyperbolicity. A detailed analysis of the hyperbolicity of the model can be found there. 

In our SOH model the order parameter $c_1(d)$ is a constant, since $d$ is a constant. Therefore, it is not able to account for the phase transitions reported e.g. in \cite{Vicsek_etal_PRL95}. In this work, the authors observe that the discrete particle model (\ref{eq:Vic_disc_X})-(\ref{eq:Vic_disc_barV}) undergoes a phase transition from disorder (small values of $c_1$) to alignment (values of $c_1$ close to $1$) as either the noise intensity $d$ is decreased or the density $\rho$ is increased. In \cite{Chate_etal_PRE08}, it is observed that in some regimes an attractor consisting of travelling bands emerges. In these bands the particle density and order parameter are large, indicating a high level of alignment. The bands travel through an ocean of low density disordered particles. Therefore, the disordered and ordered phase may coexist and are separated by dynamic interfaces. Our SOH model with a constant order parameter is unable to account for such features. 

A major step forward to resolving this inaccuracy has been made in a series of works \cite{Degond_etal_JNonlinearSci13, Degond_etal_note_submitted13, Degond_etal_preprint13, Frouvelle_Liu_SIMA12}. In \cite{Degond_etal_JNonlinearSci13, Frouvelle_Liu_SIMA12}, the collision frequency $\nu$ depends linearly on the norm of the local current i.e. $\nu = \nu_0 |{\mathcal J}_f|$, where $\nu_0$ is a constant. In this case, it is shown that, besides the VMF equilibrium, a second equilibrium coexists which consists of the uniform distribution of orientations. Additionally, at low densities, this is the only stable equilibrium. But beyond a critical density $\rho_c$, the VMF equilibria emerge as stable equilibria, while the uniform equilibria become unstable. Therefore, a second order (or continuous) phase transition happens at the critical density $\rho_c$. In \cite{Degond_etal_note_submitted13, Degond_etal_preprint13}, more general dependences of $\nu$ and $d$ upon $|{\mathcal J}_f|$ are considered and a wealth of interesting behavior, such as first order (discontinuous) phase transitions with hysteresis are exhibited. Additionally, the analysis of \cite{Degond_etal_JNonlinearSci13, Frouvelle_Liu_SIMA12} is complemented in terms of rigorous rates of convergence of equilibria and properties of the macroscopic model. In these works, it is shown that, in the area where the density is below the critical density and where consequently the isotropic equilibria are stable, the macroscopic dynamics is given by a nonlinear diffusion equation in place of the SOH model. The SOH dynamics is recovered above the critical density where the VMF distribution exists and is stable. These works pave the way to a rigorous macroscopic modeling of the patterns observed in  \cite{Chate_etal_PRE08, Vicsek_etal_PRL95}. However, to be effective, the theory should provide a way to interface the nonlinear diffusion model and the SOH model across the interface between the phases. This theory is still missing. 

The Vicsek model has been criticized on the basis that the assumption of constant speed of motion is unrealistic. It has been argued that in practice, the agent speeds in fish schools for instance is not constant. In \cite{Barbaro_Degond_DCDSB13}, we consider a model with no such constraint on the particle velocities. It consists of the Cucker-Smale (CS) model with the addition of a self-propulsion force. The CS model \cite{Carrillo_etal_SIMA10, Cucker_Mordecki_JMPA08, Cucker_Smale_IEEETransAutCont07, Ha_Liu_CMS09, Ha_Tadmor_KRM08, Motsch_Tadmor_JSP11, Shen_SIAP07} has received a lot of attention recently. Like the Vicsek model, it describes the relaxation of the agents' velocity to that of their neighbors, but there is no constraint on the agents' speed. Most previous work on the CS model ignore the self-propulsion velocity. In \cite{Barbaro_Degond_DCDSB13} we add the self-propulsion force in the form of an operator whose effect is to relax the norm of the velocity to a constant value. This operator takes the form of a Ginzburg-Landau type term in the mean-field equation. We perform the successive hydrodynamic limit and limit of large self-propulsion force in this model. We show the emergence of phase transitions between a nonlinear diffusive regime and a hyperbolic regime described by the SOH model. Therefore, the essence of the SOH model lies in the presence of a large propulsion force rather than on the normalization of the particle velocities. The importance of the self-propulsion force in triggering the phase transition in self-propelled particle systems was already remarked in~\cite{Toner_etal_AnnPhys05}. 

The SOH model also emerges as the hydrodynamic limit of a large class of microscopic dynamics. For instance, in \cite{Degond_Motsch_JSP11}, it describes the hydrodynamic limit of a system of agents controlling their motion by acting on the curvatures of their trajectories (like a driver controls the trajectory of his car by acting on the steering wheel) and trying to join their neighbor's direction of motion. The Individual model, now referred to as the Persistent Turner (PT), has been derived from experimental observations of fish trajectories in \cite{Gautrais_etal_JMB09}. The validation of the model has been made by comparing the diffusion constant induced by the model \cite{Degond_Motsch_JSP08} to the experimentally observed one \cite{Gautrais_etal_JMB09}. In \cite{Gautrais_etal_PlosCB12}, a model close to the one proposed in \cite{Degond_Motsch_JSP11} has been validated by comparisons to experimental data. 

In \cite{Degond_Liu_M3AS12}, an extension of the time-continuous particle model has been proposed in dimension $m=3$. In this model, the interaction force is complemented with a precession term. In addition to driving the agent's velocities towards their neighbors' average velocity, it makes the former rotate about the latter. The resulting hydrodynamic model is an extended SOH model with extra transport and diffusion terms acting in the direction normal to the average velocity. Interestingly enough, when the self-propulsion speed is set to $0$, the resulting SOH model is nothing but the Landau-Lifschitz-Gilbert (LLG) equations of micromagnetism \cite{Brown_Wiley63}. Therefore, our theory offers one of the very few (if not the only) derivation of the LLG equations from first principles. 

These two examples confirm that the SOH model is a generic model which may apply to a large class of situations. Additionally, in \cite{Degond_etal_preprintNash_2013}, the SOH Model provides an example of hydrodynamic models which can be derived from the interplay of kinetic theory and game theory.

\setcounter{equation}{0}
\section{Conclusion}
\label{sec_conclusion}

In this work, we have reviewed the derivation and properties of the Self-Organized Hydrodynamic (SOH) model. It describes the large-scale behavior of systems of self-propelled particles interacting through local alignment. The main difficulty in this derivation is related to the lack of conservation properties of the underlying particle dynamics. We have described how the use of the Generalized Collision Invariants can overcome this problem. The known mathematical properties of the model have been reviewed and its connections with other kinds of self-propelled particle systems have been discussed. In the future, the spatial dynamics of the interface between the SOH and nonlinear diffusion models in the case of phase transitions between ordered and disordered states will be investigated. Other kinds of constrained dynamics or geometries will be considered. The question of the boundary conditions applying to the SOH model will also be addressed.


\end{document}